\documentclass[12pt,preprint]{emulateapj}
\usepackage{graphicx}
\usepackage{url}

\shorttitle{Ellipsoidal Variations and Visual Orbit of $o$ Draconis}
\shortauthors{Roettenbacher et al.}
\slugcomment{Accepted for publication in the Astrophysical Journal}

\begin{document}

\title{Detecting the Companions and Ellipsoidal Variations of RS CVn Primaries:  \\ II.  \lowercase{$o$}~Draconis, a Candidate for Recent Low-Mass Companion Ingestion}
\author{
  Rachael M.\ Roettenbacher$^1$, John D.\ Monnier$^1$,  Francis C.\ Fekel$^2$, Gregory W.\ Henry$^2$, Heidi Korhonen$^3$, David W.\ Latham$^4$, Matthew~W.~Muterspaugh$^2$, Michael~H.~Williamson$^2$,  Fabien Baron$^1$, Theo~A.~ten~Brummelaar$^5$, Xiao~Che$^1$, Robert~O.~Harmon$^6$, Gail~H.~Schaefer$^5$, Nicholas~J.~Scott$^5$,  Judit~Sturmann$^5$, Laszlo Sturmann$^5$, and Nils~H.~Turner$^5$
  }
  \affil{$^1$Department of Astronomy, University of Michigan, Ann Arbor, MI 48109, USA \\  
  $^2$Center of Excellence in Information Systems, Tennessee State University, Nashville, TN 37209, USA \\ 
  $^3$Finnish Centre for Astronomy with ESO (FINCA), University of Turku, V\"ais\"al\"antie 20, FI-21500 Piikki\"o, Finland\\
  $^4$Harvard-Smithsonian Center for Astrophysics, 60 Garden Street, Cambridge, MA 02138, USA \\
 $^5$Center for High Angular Resolution Astronomy, Georgia State University, Mount Wilson, CA 91023, USA \\ 
  $^6$  Department of Physics and Astronomy, Ohio Wesleyan University, Delaware, OH 43015, USA 
  }
\email{Contact email:  rmroett@umich.edu}

\begin{abstract}

To measure the stellar and orbital properties of the metal-poor RS CVn binary $o$~Draconis ($o$~Dra), we directly detect the companion using interferometric observations obtained with the Michigan InfraRed Combiner at Georgia State University's Center for High Angular Resolution Astronomy (CHARA) Array.  The $H$-band flux ratio between the primary and secondary stars is the highest confirmed flux ratio ($370 \pm 40$) observed with long-baseline optical interferometry.  These detections are combined with radial velocity data of both the primary and secondary stars, including new data obtained with the Tillinghast Reflector Echelle Spectrograph on the Tillinghast Reflector at the Fred Lawrence Whipple Observatory and the 2-m Tennessee State University Automated Spectroscopic Telescope at Fairborn Observatory.  We determine an orbit from which we find model-independent masses and ages of the components ($M_\mathrm{A} = 1.35 \pm 0.05 \ M_\odot$, $M_\mathrm{B} = 0.99 \pm 0.02 \ M_\odot$, system age $=3.0\mp0.5$~Gyr).   An average of a $23$-year light curve of $o$~Dra from the Tennessee State University Automated Photometric Telescope folded over the orbital period newly reveals eclipses and the quasi-sinusoidal signature of ellipsoidal variations.  The modeled light curve for our system's stellar and orbital parameters confirm these ellipsoidal variations due to the primary star partially filling its Roche lobe potential, suggesting most of the photometric variations are not due to stellar activity (starspots).  Measuring gravity darkening from the average light curve gives a best-fit of $\beta = 0.07\pm0.03$, a value consistent with conventional theory for convective envelope stars.  The primary star also exhibits an anomalously short rotation period, which, when taken with other system parameters, suggests the star likely engulfed a low-mass companion that had recently spun-up the star.

\end{abstract}



\section{Introduction}

RS~Canum~Venaticorum (RS~CVn) stars are active binary systems with Ca~H~and~K and photometric variability \citep{hal76}.  The components of these close binary systems are typically an evolved giant or subgiant primary star with a subgiant or main sequence companion \citep{ber05, str09}. 

In close binary systems, the stars experience changes in energy and angular momentum, circularizing the orbit due to forces from the aspherical mass distributions of the (partially) Roche-lobe-filling primary stars \citep{zah77}.  Observations of open clusters have shown a transition period, the range of orbital periods between the longest circular orbit and the shortest eccentric orbit, which monotonically increases with cluster age \citep[][and references therein]{maz08}.  
Observations and predictions have shown that stars with orbital periods $P_\mathrm{orb} \lesssim 10$~days will have circularized while on the main sequence, regardless of spectral type \citep[e.g.,][]{wal49, koc81, zah89}.  As stars evolve off of the main sequence and cool while becoming subgiants and giants, circularization is expected for stars with periods $P_\mathrm{orb}\lesssim100$~days \citep[e.g.,][]{may84, cla09}.  Therefore, RS CVn systems with orbital periods longer than 100~days are likely to retain primordial non-zero eccentricities.  

The single-lined RS CVn system $o$~Draconis \citep[omicron~Draconis, $o$~Dra, 47~Dra, HD~175306, HIP~92512, HR~7125; G9III; ][]{her55,you77,wal85,str89} has a previously-measured orbital period of $P_\mathrm{orb} \sim138$ days with a non-zero eccentricity of $e \sim 0.1$ \citep[e.g.,][]{you21,luc71,mas08}.  
The primary star of $o$~Dra rotates faster than would be expected from tidal synchronization \citep[e.g.][]{gle88,mas08}.  

To achieve this rapid rotation, the primary star must have increased its angular momentum.  One possibility is that the primary star engulfed a nearby companion, an event that would have had a strong impact on the primary star and its subsequent evolution.  This evolution could shed light on the future of the Solar System as the Sun expands to ingest planets \citep[e.g., ][]{sch08}.  To investigate this hypothesis, we aim to better determine $o$~Dra's stellar parameters \citep[e.g., ][]{bas85,gur95} and thereby resolve the binary system's history.  

To advance our understanding of $o$~Dra, we present the first detections of the low luminosity companion with six nights of interferometric data and the companion's first radial velocity detections.  We describe these data sets along with radial velocity and photometric data of the primary star in Section 2.  We detail the analysis of observations with resultant orbital parameters in Section 3.  We describe the analysis of our photometry, show the detected ellipsoidal variations and eclipses, and measure gravity darkening of the primary component in Section 4.  We show our results on an Hertzsprung-Russell (H-R) diagram and discuss the system's evolution in Section 5.  Finally, we present the conclusions of our study of $o$~Dra in Section 6.  


\section{Observations}

\subsection{Interferometry}

We obtained interferometric data at Georgia State University's Center for High-Angular Resolution Astronomy (CHARA) Array.  The CHARA Array is a Y-shaped array of six $1$-m class telescopes with non-redundant baselines varying from $34$- to $331$-m located at Mount Wilson Observatory, California \citep{ten05}.  Using all six telescopes and the Michigan InfraRed Combiner \citep[MIRC;][]{mon04, mon06,che11}, we obtained $H$-band ($1.5-1.8 \ \mu$m) data (eight channels across the photometric band with $\lambda/\Delta \lambda \sim 40$) on 
UT 2012 May 9, 11, 12 and June 6, 8, 17, 18; 2014 May 25, 26, 27, June 29, 30, and July 1.  

We detected the faint companion in the data from UT 2012 May 9, 12; 2012 June 17, 18; and 2014 May 26, 27.  The  nights of observation without detections of the companion had insufficient $uv$ coverage due to poor seeing, short observation lengths, or observations during eclipse leaving the companion undetected.  We reduced and calibrated these data with the standard MIRC pipeline \citep[see][for pipeline details]{mon07,zha09,mon12}.  We used at least two calibration stars for each night of data (see Table \ref{cal}).  

Three of our seven calibrators are A stars, which were revealed to be oblate due to rapid rotation during our analysis and required more information for calibration.  Each of the stars were calibrated with a non-oblate star when possible, but some nights of observation required the use of other oblate stars.  HD 185395 ($\theta$ Cyg) was used on 2012 Jun 17 to calibrate HD 192696 (33 Cyg) and HD 184006 ($\iota$ Cyg).  On nights when HD 185395 was not available (2012 Jun 8 and 18), HD 192696 and HD 184006 were used to calibrate each other.  HD 192696 was used to calibrate HD 106591 ($\delta$ UMa) on 2012 May 9, 11, and 12).  The calibrated visibilities were fit to models of oblate stars to determine the mean uniform disk diameter, major-to-minor axis ratio, and position angle of the major axis (east of north), the mean values of which are in the footnotes of Table \ref{cal}.  

\begin{deluxetable*}{l c c c}
\tabletypesize{\scriptsize}
\tablecaption{Calibrators for $o$~Draconis}
\tablewidth{0pt}
\tablehead{
\colhead{Calibrator Name} & \colhead{Calibrator Size (mas)$^a$} & \colhead{Source} & \colhead{UT Date of Observation}
}
\startdata
HD 106591 ($\delta$ UMa) & oblate$^b$ & MIRC calibration & 2012 May 9, 12\\
HD 125161 ($\iota$ Boo) &  $ 0.49 \pm 0.03 $ & \cite{bon06} &  2012 May 9, 12 \\
HD 138852 & $ 1.00 \pm 0.06 $ & \cite{bon06} & 2014 May 26 \\
HD 184006 ($\iota$ Cyg) & oblate$^c$ & MIRC calibration & 2012 Jun 17, 18; 2014 May 26, 27 \\
HD 185264 & $ 0.80 \pm 0.06 $ & \cite{bon06} & 2014 May 26, 27 \\
HD 185395 ($\theta$ Cyg) & $ 0.73 \pm 0.02 $ & \cite{whi13} & 2012 Jun 17\\
HD 192696 (33 Cyg) & oblate$^d$ & MIRC calibration & 2012 May 9, 12, Jun 17, 18
\enddata
\tablecomments{$^a$ Some calibration stars are rapidly rotating and distorted from spherical.  They were modeled as $H$-band uniform ellipses for calibration.  For each star we list the $H$-band uniform disk mean diameter (mas), major-to-minor axis ratio, and position angle of the major axis ($^\circ$, east of north). \\
$^b$ $\theta_\mathrm{UD, mean} = 0.78\pm 0.02$, major/minor $= 1.29 \pm 0.05$, PA$_\mathrm{maj} = 113 \pm 5^\circ$ \\
$^c$ $\theta_\mathrm{UD, mean} = 0.72\pm 0.02$, major/minor $= 1.29 \pm 0.04$, PA$_\mathrm{maj} = 92 \pm 4^\circ$ \\
$^d$ $\theta_\mathrm{UD, mean} = 0.56\pm 0.02$, major/minor $= 1.41 \pm 0.09$, PA$_\mathrm{maj} = 115 \pm 6^\circ$}
\label{cal}
\end{deluxetable*}

\subsection{Radial Velocity}

We combined three radial velocity data sets of the primary star and one set for the secondary to further constrain our orbit of $o$~Dra. 

Three radial velocities published in \citet{mas08} were obtained with the CfA Digital Speedometer on the Wyeth Reflector at the Oak Ridge Observatory (Harvard, MA; 2004 September 1$-$ 2005 April 14).  Two radial velocities also published in \citet{mas08} were obtained with the CfA Digital Speedometer on the Tillinghast Reflector at the Fred L.\ Whipple Observatory (Mount Hopkins, AZ; 2006 June 14 $-$ 2007 September 21).  
Six additional radial velocities (2012 May 9 $-$ 2012 October 10) for the primary star were taken with the Tillinghast Reflector Echelle Spectrograph (TRES) at the Tillinghast Reflector (see Table \ref{RVCfA}).

\begin{deluxetable}{l c}
\tabletypesize{\scriptsize}
\tablecaption{Radial Velocity Data of $o$~Draconis (Wyeth \& Tillinghast/CfA)}
\tablewidth{0pt}
\tablehead{
\colhead{HJD $-2400000$} & \colhead{Primary (km s$^{-1}$)} 
}
\startdata
    $53249.7015$ & $-27.27$ \\
    $53306.5778$ & $-31.61$ \\
    $53474.7086$ &  $4.10$ \\
    $53900.9303$  &  $2.21$ \\
    $54364.6628$ & $-31.30$ \\
    $56056.9907$ & $-42.65$ \\
    $56075.9832$ & $-30.30$ \\
    $56110.9890$  &  $3.86$ \\
    $56132.6637$  & $-6.89$ \\
    $56198.7292$ & $-41.88$ \\
    $56210.5694$ & $-34.50$ 
\enddata
\tablecomments{Residuals on the primary radial velocities are $0.39$ km s$^{-1}$.  The first five velocities are published in \citet{mas08}.} 
\label{RVCfA}
\end{deluxetable}

With the Tennessee State University 2-m automatic spectroscopic telescope (AST) at Fairborn Observatory, AZ, we have determined radial velocities from 86 spectrograms of the primary of
$o$~Dra taken between 2007 October 11 $-$ 2014 October 28 and 19 measurements of the companion radial velocity  (2012 October 10 $-$ 2014 October 28; see Table \ref{RVTSU}).  See Appendix A, \citet{fek09}, and \citet{roe15} for details on these observations.

\begin{deluxetable}{l c c}
\tabletypesize{\scriptsize}
\tablecaption{Radial Velocity Data of $o$~Draconis (AST/TSU)}
\tablewidth{0pt}
\tablehead{
\colhead{HJD $-2400000$} & \colhead{Primary (km s$^{-1}$)} & \colhead{Secondary (km s$^{-1}$)} 
}
\startdata
    $54384.710$  &  $-41.9$ & \\
    $56210.620$   & $-34.5$ & $-1.2$ \\
    $56211.589$    &$-33.4$ & \\
    $56212.589$  &  $-32.6$ & \\
    $56227.713$   & $-12.2$ & 
\enddata
\tablecomments{Table \ref{RVTSU} is published in its entirety in the electronic edition.  Residuals on the primary radial velocities are $0.22$ km s$^{-1}$ and secondary are $1.6$ km s$^{-1}$. } 
\label{RVTSU}
\end{deluxetable}

We additionally include 18 archival radial velocities of the primary star from \citet[][1902 July 15 $-$ 1920 August 9]{you21}.

\subsection{Photometry}

\begin{figure}
\hspace{-0.3cm}
\includegraphics[angle=90,scale=0.38]{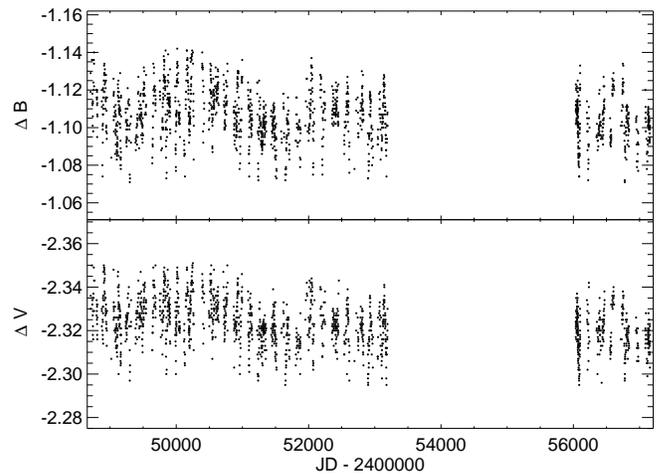}
\caption{Johnson $B$ and $V$ differential magnitudes of $o$~Dra acquired over 23 years from 1992 $-$ 2015 with the T3 $0.4$-meter APT at Fairborn Observatory in southern Arizona.}
\label{omiDraLC}
\end{figure}

\begin{figure}
\hspace{-0.3cm}
\includegraphics[angle=90,scale=0.38]{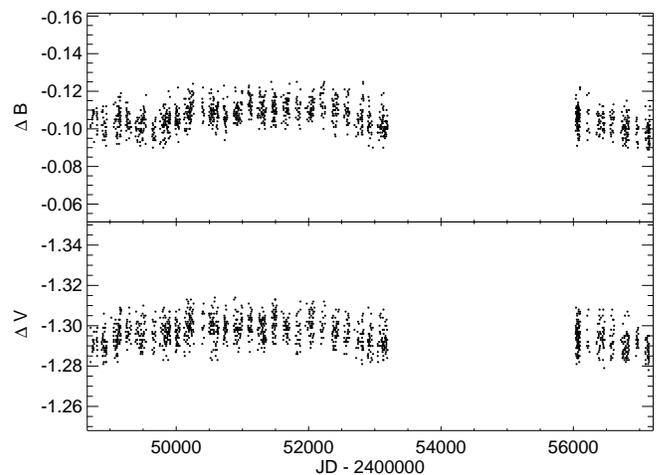}
\caption{Johnson $B$ and $V$ differential magnitudes of comparison star HD~175511 (with the check star 48~Dra).}
\label{compLC}
\end{figure}

$o$~Dra has been monitored since 1992 with Tennessee State University's T3 0.4-m Automatic Photometric Telescope (APT) at Fairborn Observatory.  Our observations span over 23 years from 1992 March 24 $-$ 2015 May 13, but with a gap during 2005$-$2011 (see Table \ref{LC} and Figure \ref{omiDraLC}).  Our Johnson \emph{B} and \emph{V} measurements of $o$~Dra were made differentially with respect to the comparison star HD~175511 (HIP~92594; B9; see Figure \ref{compLC}) and the check star HD~176408 (K1III).  Details of the robotic telescopes and photometers, observing procedures, and data analysis techniques can be found in \citet{hen99} and \citet{fek05}.  

\begin{deluxetable*}{l c c c c}
\tabletypesize{\scriptsize}
\tablecaption{Differential Johnson B and V Photometric Data of $o$~Draconis (APT/TSU)}
\tablewidth{0pt}
\tablehead{
\colhead{HJD$- 2400000$} & \colhead{$\Delta B$ ($o$ Dra $-$} & \colhead{$\Delta V$ ($o$ Dra $-$ } & \colhead{$\Delta B$ (48 Dra$^a$ $-$ } & \colhead{$\Delta V$ (48 Dra $-$ } \\
& \colhead{ HD 175511)} & \colhead{HD 175511)} & \colhead{HD 175511)} & \colhead{HD 175511)}
}
\startdata
$48705.9454$        &    $-1.119$        &    $-2.330$     & $-0.101$ & $-1.282$ \\
$48706.9415$        &    $-1.129$        &    $-2.336$     & $-0.102$ & $-1.299$ \\
$48725.8916$        &    $-1.136$        &    $-2.350$     &   &\\
$48732.8732$        &    $-1.113$        &    $-2.332$     & $-0.106$ & \\
$48733.8831$        &    $-1.118$        &    $-2.334$     & $-0.105$ & $-1.287$
\enddata
\tablecomments{Table \ref{LC} is published in its entirety in the electronic edition.  \\
$^a$ HD~175511 (HIP~92594), $V = 6.94$, $B-V = -0.04$. \\
$^b$ 48~Dra (HD~176408, HIP~92997), $V = 5.68$, $B-V = 1.16$. }
 
\label{LC}
\end{deluxetable*}

\subsection{High-Resolution Spectroscopy}

We use high-resolution spectra of $o$~Dra covering the spectral range of $3700-7300$~\AA \ obtained at the Nordic Optical Telescope (NOT) with the FIES high resolution echelle spectrograph. In the current work, the $1.3$~arcsec fiber giving a resolving power ($\lambda/\Delta\lambda$) of $67000$ was used. The exposure time for each spectrum was $180$~s and resulted in a typical signal-to-noise ratio of $500$ per resolution element (four pixels) at $6420$~\AA.  
The observations were carried out at thirteen epochs between 2012 April 17 $-$ August 15.  All the spectra were reduced with the dedicated FIES reduction software FIEStool.


\section{Orbital Elements and Masses}

Using model fitting for the location of the unresolved secondary star with respect to the resolved primary star, we directly detected the companion of $o$~Dra.  Our models allow the primary star's major axis and major-to-minor axis ratio, primary-to-secondary $H$-band flux ratio, and the secondary's position to vary.  During the fitting for the companion we weighted the closure phases ten times more strongly in the final $\chi^2$ than the squared visibilities and triple amplitudes in order to better identify asymmetries in the system to detect the faint companion  (the detection of the companion is not sensitive to the factor of 10; see Figures \ref{omiDraVis} $-$ \ref{omiDraT3} in Appendix B).  Errors for the primary star's size and shape and the primary-to-secondary $H$-band flux ratio are based on the individual epochs.  The positional errors of the location of the secondary are error ellipses based upon the shape of the $\chi^2$ surface used to detect the companion.

The coordinates of the secondary detections on six nights (UT 2012 May 9, 12; June 17, 18; 2014 May 26 and 27) are listed in Table \ref{detect}.  We measured the $H$-band uniform disk diameter of the primary star to be $\theta_\mathrm{UD} = 2.115\pm0.007$ mas (limb-darkened disk $\theta_\mathrm{LD} = 2.189\pm0.007$, obtained by assuming a limb darkening power law exponent $\alpha = 0.27$) with a major-to-minor axis ratio of $1.01\pm0.03$.  The $H$-band flux ratio for the primary star to the secondary is $370\pm40$, the highest confirmed binary flux ratio measured with long-baseline optical interferometry \citep[RT Aur has an unconfirmed $H$-band flux ratio of $\sim450$:1, measured by MIRC at the CHARA Array; ][]{gal15}.

\begin{deluxetable*}{l c c c c c c c}
\tabletypesize{\scriptsize}
\tablecaption{Detections for the companion of $o$~Draconis with respect to the primary}
\tablewidth{0pt}
\tablehead{
\colhead{UT Date} & \colhead{JD $- 2400000$} & \colhead{Separation} & \colhead{Position} & \colhead{Error Ellipse} & \colhead{Error Ellipse} & \colhead{Error Ellipse} & \colhead{Reduced} \\
 & &  \colhead{(mas)} & \colhead{ ($^\circ$)$^a$} & \colhead{Major Axis (mas)$^b$} & \colhead{Minor Axis (mas)$^b$} & \colhead{Position Angle ($^\circ$)$^a$} & \colhead{$\chi^2$}
 } 
\startdata
2012 May 09 & 56056.97 & 6.52  & 23.1 & 0.06 & 0.04 & 340 & 1.3\\
2012 May 12 & 56059.95 & 6.36  & 23.7 & 0.05 & 0.03 & 60 & 1.5\\
2012 June 17 & 56095.87 & 3.71 &  203.1 & 0.04 & 0.03 & 310 & 1.6\\
2012 June 18 & 56096.82 & 3.96 &  202.3 & 0.02 & 0.02 & 330 & 1.5\\ 
2014 May 26 & 56803.87 & 6.03 &  202.5 & 0.07 & 0.05 & 20 & 0.9\\
2014 May 27 & 56804.85 & 6.04 &  202.0 & 0.16 & 0.04 & 330 & 1.1
\enddata
\tablecomments{These detections give an $H$-band ($1.5-1.8 \ \mu$m) primary-to-secondary flux ratio for $o$~Dra A to B of $370\pm40$.  The uniform disk fit for the primary star is $2.115\pm0.007$ mas (limb-darkened disk $2.189\pm0.007$) with a $1.01\pm0.03$ major-to-minor axis ratio. \\
$^a$East of North\\
$^b$Scaled error bars to ensure a reduced $\chi^2 = 1.00$ as described in Section 3. } 
\label{detect}
\end{deluxetable*}

We calculate the orbital parameters of the binary by simultaneously fitting our interferometric and radial velocity data with Monte Carlo realizations.  For our six interferometric points we present the  error ellipses of the major and minor axis in Table \ref{detect} scaled to give our fit a reduced $\chi^2 = 1.00$.  
The radial velocity errors are similarly scaled to require $\chi^2 = 1.00$ ($\mathrm{rms}_\mathrm{CfA,A} = 0.39$~km s$^{-1}$, $\mathrm{rms}_\mathrm{AST,A} = 0.22$~km s$^{-1}$, $\mathrm{rms}_\mathrm{AST,B} = 1.6$ km s$^{-1}$, $\mathrm{rms}_\mathrm{Young, A} = 2.2$~km s$^{-1}$).

\begin{figure}
\hspace{-0.5cm}
\vspace{0cm}
\includegraphics[angle=90,scale=0.6]{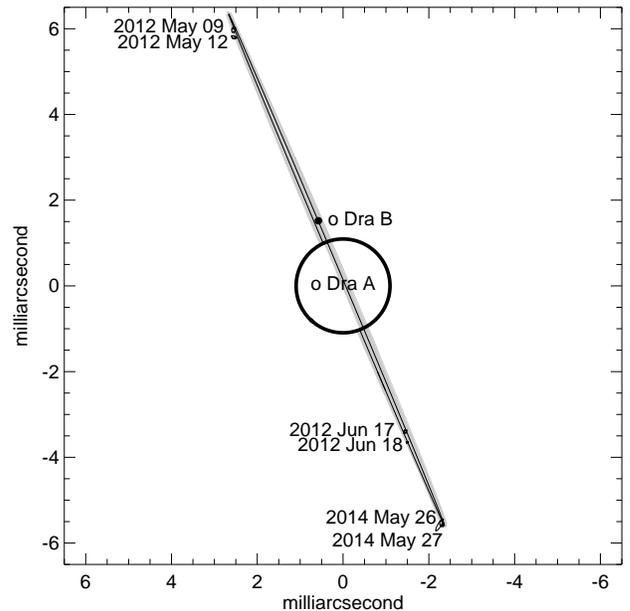}
\vspace{0cm}
\caption{Visual orbit of the RS~CVn system $o$~Dra with our dates of companion detection and their locations on the orbit (black error ellipses).  The observed stellar radius of the primary star is plotted with a thick black line ($o$~Dra~A).  The radius of the companion star, $o$~Dra~B, for the best-fit temperature of $6000$~K is plotted as the small black circle.  The light gray orbits represent fifty Monte Carlo realizations.  Black lines connect the center of the detection error ellipse to the expected point in the best-fit orbit, which is overlaid in black (given in Table \ref{omiDraparam} with $1-\sigma$ errors).  At the southernmost point in the orbit, the secondary star is moving toward the observer.  Note:  the axis units are milliarcseconds (mas) with north upwards and east to the left.}
\label{omiDradetection}
\end{figure}

\begin{figure}
\hspace{-0.5cm}
\vspace{0cm}
\includegraphics[angle=90,scale=.38]{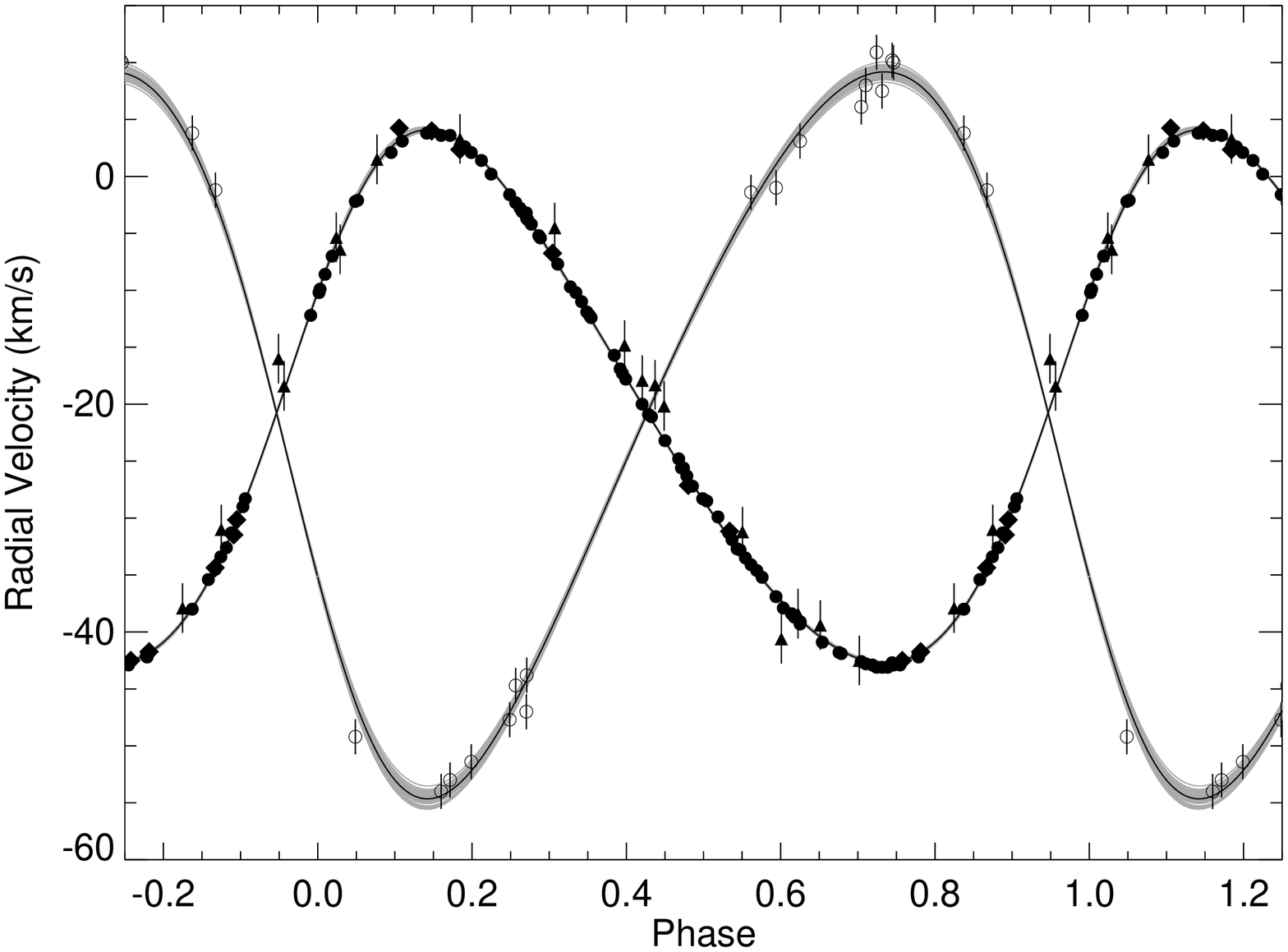}
\vspace{0cm}
\caption{Radial velocity curves of the components of $o$~Dra.  The filled symbols represent measurements of the primary star.  The filled diamonds represent observations from \citet{mas08}/CfA. The filled circles are the AST observations.  The filled triangles are observations from \citet{you21}.  $1-\sigma$ errors in velocity are plotted unless the error is smaller than the symbols.  Similarly, the open circles represent AST radial velocity data for the secondary star with $1-\sigma$ error bars.  The gray orbits are fifty Monte Carlo realizations (dark gray for the primary star and light gray for the secondary star) with the best-fit orbital parameters overlaid in black.  See Table \ref{omiDraparam} for orbital parameters with $1-\sigma$ errors.}
\label{omiDraRV}
\end{figure}

The simultaneous radial velocity and astrometry Monte Carlo realizations of the orbit gave the orbital parameters and their $1-\sigma$ errors listed in Table \ref{omiDraparam}.  The orbit is represented in Figure \ref{omiDradetection} with the radial velocity curve in Figure \ref{omiDraRV}. 
We use the conventions presented by \citet{hei78}, where the argument of periastron, $\omega$, follows the radial velocity orbit convention (the primary star with respect to the center of mass), which is different from the visual orbit convention (the secondary star with respect to the primary).  The position angle of the  ascending node (E of N), $\Omega$, is independent of definition being equivalent with respect to either the primary or secondary star.

\begin{deluxetable*}{l c}
\tabletypesize{\scriptsize}
\tablecaption{Orbital and Stellar Parameters of $o$~Draconis}
\tablewidth{0pt}
\tablehead{
\colhead{Measured Parameters} & \colhead{Value} 
}
\startdata
semi-major axis, $a$ (mas) & $6.51\pm0.03$\\
eccentricity, $e$ & $0.158\pm0.003$\\
inclination, $i$ ($^\circ$) & $89.6\pm0.3$\\
argument of periastron, $\omega$ ($^\circ$)$^a$ & $293.0\pm0.6$\\
ascending node, $\Omega$ ($^\circ$, E of N) & $ 22.9\pm0.2$\\
period, $P_\mathrm{orb}$ (days) & $138.444\pm0.003$\\  
time of periastron passage, $T$ (HJD) & $2454983.0\pm0.2$\\  
velocity semi-amplitude, $K_\mathrm{A}$ (km s$^{-1}$) & $23.42\pm0.05$\\
velocity semi-amplitude, $K_\mathrm{B}$ (km s$^{-1}$) & $32.0\pm0.4$\\
system velocity, $\gamma$ (km s$^{-1}$) & $-20.77\pm0.04$\\
$H$-band uniform disk diameter, $\theta_{\mathrm{UD},A}$ (mas)&$2.115\pm0.007$\\  
$H$-band limb-darkened disk diameter, $\theta_{\mathrm{LD},A}$ (mas)$^b$&$2.189\pm0.007$\\  
primary major-to-minor axis ratio & $1.01 \pm 0.03$\\
$B$-band flux ratio, primary-to-secondary & $60 \pm 20$\\
$V$-band flux ratio, primary-to-secondary & $130 \pm 80$\\
$H$-band flux ratio, primary-to-secondary & $370 \pm 40$\\
rotational velocity, $v \sin i$ (km s$^{-1}$) & $16.0 \pm 0.5$\\
\hline
\colhead{Derived Parameters} & \\
\hline
orbital parallax, $\pi$ (mas) & $9.36\pm0.10$\\
distance, $d$ (pc) & $106.8 \pm 1.1$\\ 
primary radius, $R_\mathrm{A}$ ($R_\odot$)$^c$ & $25.1\pm0.3$\\ 
primary luminosity, $L_\mathrm{A}$ ($L_\odot$) & $220\pm30$\\  
primary surface gravity, $\log g_\mathrm{A}$ (cm/s$^2$) & $1.769\pm$0.007\\  
primary mass, $M_\mathrm{A}$ ($M_\odot$) & $1.35\pm0.05$\\ 
primary rotation period, $P_\mathrm{rot}$ (days)$^d$ & $79 \pm 3$\\
secondary radius, $R_\mathrm{B}$ ($R_\odot$) & $1.0 \pm 0.1$\\
secondary luminosity, $L_\mathrm{B}$ ($L_\odot$) & $1.3 \pm 0.2$\\
secondary surface gravity, $\log g_\mathrm{B}$ (cm/s$^2$) & $4.43\pm0.09$\\
secondary mass, $M_\mathrm{B}$ ($M_\odot$) & $0.99\pm0.02$\\
secondary temperature, $T_{\mathrm{eff},B}$ (K) & $6000^{+400}_{-300}$ \\
system age (Gyr) & $3.0 \mp0.5$\\    
\hline
\colhead{Literature Parameters} & \\
\hline
primary effective temperature, $T_{\mathrm{eff},A}$ (K)$^e$ & $4430\pm130$\\
primary metallicity, Fe/H$^f$ & -0.5
\enddata%
\tablecomments{$^a$Radial velocity convention for primary with respect to the center of mass.  \\ 
$^b$We applied a $3.5\%$ correction from uniform to limb-darkened disk diameter, which is consistent with a limb-darkening power law exponent of $\alpha = 0.27$.\\
$^c$Using limb-darkened disk diameter.\\
$^d$Assuming $i_\mathrm{rot} = i_\mathrm{orb}$.\\
$^e T_{\mathrm{eff},A}$ is an average of temperatures given by  \citet{chr77,gle77,rut87,gle88,mcw90,luc91,pou03,boe04,mas08,sou10,mcd12}.  The $1-\sigma$ error is the standard deviation of these values. \\
$^f$[Fe/H] is approximated in stellar evolution models based upon values given by \cite{mcw90,mas08,sou10}.} 
\label{omiDraparam}
\end{deluxetable*}

Our double-lined and visual orbit confirms previous analyses of orbital period and non-zero eccentricity \citep[e.g.,][]{you21,luc71}, while highlighting new evidence that the system is eclipsing ($i = 89.6 \pm 0.3^\circ$).  We determine the masses of the stars to be $M_\mathrm{A} = 1.35 \pm 0.05 \ M_\odot$ and $M_\mathrm{B} = 0.99 \pm 0.02 \ M_\odot$; the implications to system evolution will be discussed in Section 5.  
Our orbit gives a parallax of $\pi = 9.36\pm0.10$~mas, a value consistent with \citet[][$\pi = 9.54\pm0.21$~mas]{van07} but about $2-\sigma$ from the original Hipparcos reduction \citep[$\pi = 10.12\pm0.43$~mas; ][]{ESA97}, confirmed by \citet[][$\pi=10.27 \pm 0.42$ mas]{pou03}.  Adopting the
spectroscopic orbit presented here, the resulting Hipparcos parallax becomes
$\pi = 10.15 \pm 0.43$ mas (Pourbaix, private communication).
 The origin of the $2-\sigma$ discrepancy in parallax measurements is not well-understood; for our purposes in later sections, we proceed with our derived value of  $\pi = 9.36\pm0.10$~mas.


\section{Light Curve Models and Ellipsoidal Variations}

A previous study, \citet{str89}, suggested $o$~Dra has photometric variations up to $\Delta \ V \sim 0.01$~mag.  Other indications of activity include chromospheric Ca~H~and~K emission and variation \citep{you77, sim87, str88}.  \citet{hal86b} reported a possible rotation period of $P_\mathrm{rot} = 54.6$~days from photometric observations.  Using our NOT spectra, we can determine the value of $v \sin i$.  We modeled the Fe~$6421$~\AA \ line with a variety of $v \sin i$ models from  \citet{cas04} with  $T_\mathrm{eff,A}= 4530$~K, $\log g_\mathrm{A} = 1.77$, [Fe/H]$ = -0.5$, microturbulence $\xi_\mathrm{m} = 1.4$~km s$^{-1}$, and macroturbulence $\xi_\mathrm{M} =6$~km s$^{-1}$.  The resultant value is $v \sin i = 16 \pm 0.5$~km s$^{-1}$ when combined with our observed primary radius and orbital inclination (assuming $i_\mathrm{orb} = i_\mathrm{rot}$) gives $P_\mathrm{rot,A} = 79 \pm 3$~days.  

\begin{figure}
\hspace{-0.5cm}
\includegraphics[angle=90,scale=.38]{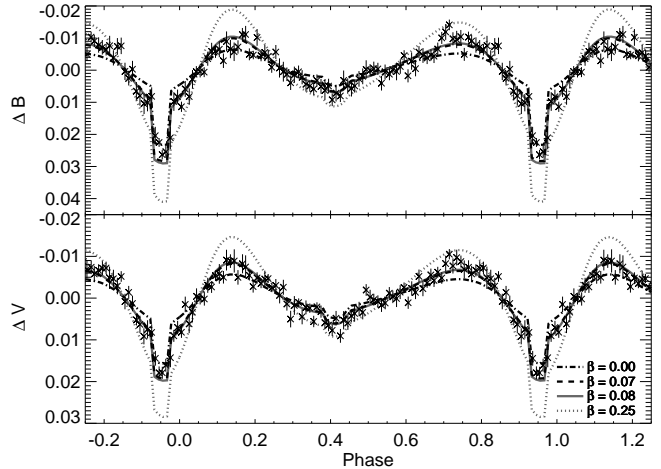}
\caption{Differential folded and binned light curves of $o$~Dra for $B$ and $V$ magnitudes plotted with error bars from the binning.  Each data point is an average of data points spanning $0.01$ in phase from the complete light curve folded on the orbital period.  The quasi-sinusoidal signature of the averaged light curve is due to ellipsoidal variations caused by the primary star partially filling its Roche lobe potential.  The lines represent the ELC models for ellipsoidal variations with the gravity darkening coefficient $\beta = 0.00$, $0.07$, $0.08$, and $0.25$, where $\beta = 0.07\pm0.03$ is the best fit to the binned light curves.  
}
\label{omiDrafit}
\end{figure}

We investigate the APT Johnson $B$ and $V$ differential light curves for evidence of starspots.   We removed long-term variations ($\Delta V \sim 0.02$) that may be attributed to axisymmetric spot structures or polar spot structures.  We folded the adjusted light curves over the orbital period ($P_\mathrm{orb} = 138.444$~days) and binned the data ($0.01$ in phase).  The resultant Johnson $B$ and $V$ light curves are presented in Figure \ref{omiDrafit}.  The quasi-sinusoidal trend observed in the averaged light curves suggests ellipsoidal variations due to distortions of the primary star partially filling its Roche lobe potential \citep[$R_\mathrm{L} = 60.8 \ R_\odot$, $R_A/R_\mathrm{L} = 0.42$; e.g., ][]{roe15}.  Both light curves clearly show an eclipse (phase $\sim 0.95$) and evidence of a weak eclipse (phase $\sim 0.40$).  Comparing the timing of the eclipses to the radial velocity curve and visual orbit, we see that the deeper eclipse is associated with the secondary star moving behind the primary star, revealing that the secondary star is hotter than the primary star.  

The secondary eclipse provides flux ratios for Johnson $B$ and $V$ bands ($60\pm20$ and $130\pm80$, respectively).  Using the $H$- and $B$-band flux ratios, we are able to constrain the temperature of the companion star.  We use NextGen stellar atmospheres \citep{hau99} restricted to the bandpasses with the size and temperature of the primary star to determine the secondary star's effective temperature ($T_\mathrm{eff,B} = 6000^{+400}_{-300}$~K), radius ($R_\mathrm{B} = 1.0 \pm0.1 \ M_\odot$), luminosity ($L_\mathrm{B} = 1.3\pm 0.2 \ L_\odot$), and surface gravity ($\log g_\mathrm{B} = 4.43\pm0.09$ (cm/s$^2$)).   

To model the observed ellipsoidal variations, we used the light-curve-fitting software package Eclipsing Light Curve \citep[ELC;][]{oro00}.  ELC accurately accounts for the star's ellipsoidal shape which changes as the companion star moves along its eccentric orbit.  We begin by modeling the light curves with no free parameters.  We assume that the orbital and rotational axes are aligned ($i_\mathrm{orb} = i_\mathrm{rot}$, $P_\mathrm{rot} = 79$~days), although our results are not sensitive to the assumed $P_\mathrm{rot}$.
We assume gravity darkening to be $\beta = 0.08$ \citep{luc67}.  The ellipsoidal variations from the modeled light curves agree remarkably well with the observed light curve of $o$~Dra confirming that  the coherent quasi-sinusoidal signature is due to ellipsoidal variations (see Figure \ref{omiDrafit}).  

\begin{figure}
\hspace{-0.5cm}
\includegraphics[angle=90,scale=.38]{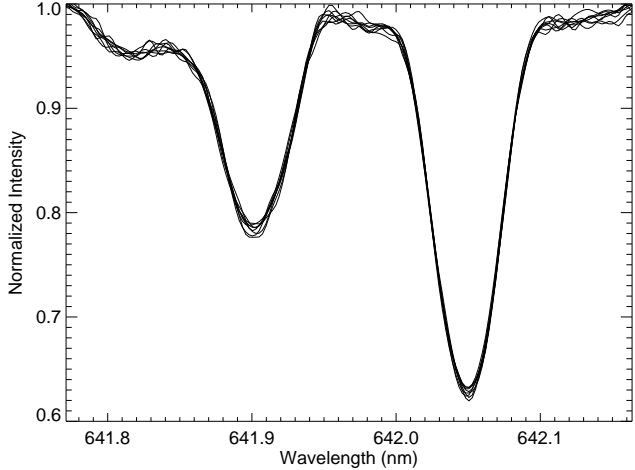}
\caption{Portion of the $o$~Dra spectrum containing two Fe I (6419 and 6421 \AA) lines.  Eight spectra are over-plotted spanning JD 2012 April 17 $-$ August 15, approximately two rotation periods.  Starspots moving across the stellar surface will manifest as features moving through some absorption lines.  While the lines deviate from Gaussian profiles, throughout the rotation of $o$~Dra, the absorption line cores do not vary in a periodic way, suggesting a featureless surface.  
}
\label{omiDraspec}
\end{figure}

The long-term variations, ellipsoidal variations, and eclipses account for much of the large changes in the light curves, suggesting that the previously identified starspots \citep[e.g.,][]{hal86b, str89} were instead observations of a combination of these effects.  Additionally, the absorption lines of the NOT spectra do not show clear evidence of rotationally-modulated temperatures due to starspots (see Figure \ref{omiDraspec}).  

\begin{figure}
\hspace{-0.5cm}
\includegraphics[angle=90,scale=.38]{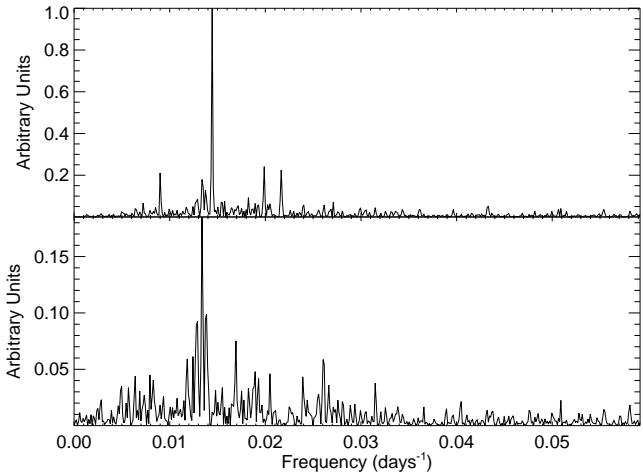}
\caption{Power spectrum of the Johnson $B$ light curve of $o$~Dra.  In the top panel, power spectrum of the $B$ light curve with the long-term trend removed shows a peak at half the orbital period ($69.2$~days), consistent with the signature of ellipsoidal variations.  In the bottom panel, the power spectrum of the $B$ light curve with the long-term trend, eclipses, and ellipsoidal variations removed shows a peak at $75$ days.  
}
\label{omiDraFFT}
\end{figure}

However, weak spot signatures are occasionally still visually present after removing the eclipse and ellipsoidal variation model from the observed light curve.  To determine a rotational period based on the spot signature, we perform a power spectrum analysis in which we sampled the light curve on a grid of one-day spacing (inserting 0.0 on days without data).  We found the strongest signature comes from a period of $P_\mathrm{rot} = 75$~days (see Figure \ref{omiDraFFT}), slightly smaller than the estimated $79$-day period based upon $v \sin i$.  While $P_\mathrm{rot} = 75$~days is within $2\sigma$ from the spectroscopically determined rotation period, this small difference could be attributed to differential rotation, often seen with RS CVn stars \citep{str09}.

Because the starspots are relatively weak, they do not strongly contaminate the phase-averaged light curve of $o$~Dra, providing an opportunity for precision analysis.
Using ELC, we measured the level of gravity darkening by modeling the system with no free parameters except the gravity darkening coefficient, $\beta$ from $T_\mathrm{eff} \propto g^\beta$ \citep{von24}.  We found that the best-fit gravity darkening coefficient for $o$~Dra is $\beta = 0.07 \pm0.03$ (errors from bootstrapping over observational seasons), similar to recent findings of \citet[][$\beta = 0.06 \pm 0.01$]{dju06}
and \citet[][$\beta<0.1$]{roe15}.  These measurements are consistent with the canonical value $\beta \sim 0.08$ from \citet{luc67}, but differs from an alternative value $\beta \sim 0.21$ from \citet{esp12}.

\section{Hertzsprung-Russell Diagram and Evolutionary History}

\begin{figure}
\hspace{-0.3cm}
\vspace{0cm}
\includegraphics[angle=90,scale=0.37]{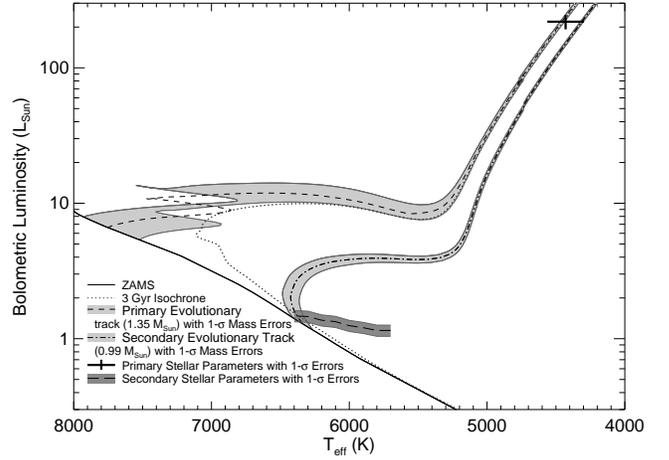}
\vspace{0cm}
\caption{H-R diagram for $o$~Dra.  The dashed and dot-dashed lines are the main sequence and post-main sequence evolutionary tracks for $1.35 \ M_\odot$ and $0.99 \ M_\odot$ stars with $[Fe/H]\sim-0.5$, respectively \citep{dot08}.  The gray regions represent our $1-\sigma$ mass errors with the solid black line representing the zero age main sequence.  The dotted line is a $3$~Gyr isochrone \citep[PHOENIX; ][]{dot08}.  The measured location of the primary with $1-\sigma$ errors is indicated by the plus sign.  The companion location is indicated with the long-dashed line (with $1-\sigma$ errors in luminosity).}
\label{omiDraHR}
\end{figure}

To understand why $o$~Dra has $P_\mathrm{rot} < P_\mathrm{orb}$, we investigate the evolution of the present stellar components.  We plot the location of the components of $o$~Dra on an H-R diagram (see Figure \ref{omiDraHR}) using our measured stellar parameters determined from our orbit and flux ratios.  We include the zero-age main sequence and  Dartmouth stellar evolutionary tracks \citep[Fe/H $=-0.5$, $\alpha$/Fe $=0.0$, PHOENIX-based models,][]{dot08} for interpolated model masses ($M_{A,\mathrm{model}} = 1.35 \pm 0.05 \ M_\odot$,  $M_{B,\mathrm{model}} = 0.99 \pm 0.02 \ M_\odot$).  

Our primary star detection falls on the $1.35 \ M_\odot$ evolutionary track with an estimated temperature of $4430 \pm 130$ K \citep{chr77,gle77,rut87,gle88,mcw90,luc91,pou03,boe04,mas08,sou10,mcd12}.  The detection of the secondary with $1-\sigma$ errors cross the main sequence of the expected evolutionary track, but at the upper limit of our temperature range ($T_{\mathrm{eff},B} = 6000^{+400}_{-300}$~K).  Our system age estimate most strongly depends upon the mass of the primary star; for the primary mass ($M_{A,\mathrm{model}} = 1.35 \pm 0.05 \ M_\odot$), we determine an age of the system of $3.0 \mp0.5$~Gyr.  

We can use our knowledge of the evolutionary state of $o$~Dra to investigate three possible explanations why the rotational period is faster than its orbital period. 
First, we conclude that the star could not have evolved right off of the main sequence.  The rotational velocity is more rapid than expected ($\sim 3$~km s$^{-1}$) for the evolution of a $1.35 \ M_\odot$, early F main sequence star \citep[based upon initial rotational periods of $\sim 2$ days and main-sequence radius of $\sim 2 \ R_\odot$;][]{nie13,boy12}.  Another component of the system must have imparted some angular momentum.  

Second, secondary stars can spin-up primary stars faster than the orbital period in the case of eccentric orbits.  Pseudosynchronous orbits are those that have the rotation of the star synchronized with the periastron passage of the secondary star \citep{hut81}.  $o$~Dra has a predicted pseudosynchronous rotation period of $P_\mathrm{ps} = 123.3 \pm 2.3$~days \citep{hal86}, which is much longer than our observed rotational period.  Thus, the known companion of $o$~Dra could not have spun up the system to its current rotational period.  

Third, a now-unseen companion could have spun-up the primary star.  Because the secondary star has a mass of $M_\mathrm{B} = 0.99 \pm 0.02 \ M_\odot$, the two objects that merged would have to have a significant difference in mass, otherwise the secondary star would be more evolved than the primary.  With the primary and secondary stars falling on the same isochrone, the component that merged into the primary must have been only a small fraction of primary mass so as to not significantly affect its evolution.  
In order be consistent with rotational velocity observations, sufficient angular momentum would need to be imparted by the merging component to spin-up the $\sim 1.04 \ M_\odot$ convective envelope of the $1.35 \ M_\odot$ primary star\footnote{The mass of the convective envelope was determined using EZ-Web, 
\url{http://www.astro.wisc.edu/$\sim$townsend/static.php?ref=ez-web}, R.\ H.\ D.\ Townsend's Web-browser interface of the Evolve ZAMS evolution code \citep{pax04}}.  Unfortunately, the rotation period of the star when the companion was engulfed is unknown and the primary star could have since slowed.  While we cannot accurately estimate the mass of the consumed companion, we find from angular momentum arguments that the companion could range in mass from a giant planet to a low-mass star.  
If a low-mass companion were initially present in the system at $0.1$~AU, it would be dynamically stable with the system's stellar components on the timescale of the system age, $3$~Gyr, but would be engulfed by the primary star as the evolving stellar radius approached the companion's orbital radius \citep[cf.\ ][]{dav03}.


\section{Conclusions}

In this work, we have made the first visual detections of the secondary star of $o$~Dra using interferometric observations.  The $H$-band primary-to-secondary flux ratio is $370\pm40$, the highest confirmed flux ratio for a binary detected with long-baseline optical interferometry.  With the astrometry and radial velocity data for both stars, we establish the first full three-dimensional orbit to determine orbital and stellar parameters.

By folding and binning photometric data, we have shown evidence of ellipsoidal variations, gravitational distortions of the primary star caused by the close companion.  The observed light curves are nearly identical to light curve models generated only from stellar and orbital parameters leading to the conclusion that the primary star has ellipsoidal variations, as opposed to long-lived starspots or active longitudes.  After removing the model light curve, we observe only weak signs of rotationally-modulated starspots.  Additionally, there could be potentially active regions (e.g., axisymmetric spot structures and polar spots) that affect the stellar flux over longer periods of time with a global brightening or dimming.  The folded, binned light curve also shows primary and secondary eclipses, which provide flux ratios to help constrain the stellar parameters of the secondary star.

Our new, high-precision orbital elements along with the folded light curve also allow for a measurement of gravity darkening.  We find that $\beta = 0.07\pm0.03$, a value of gravity darkening consistent with conventional theory \citep{luc67} and previous results \citep{dju06,roe15}.  

We have established that the primary star's rapid rotation period could be due to the transfer of angular momentum from a nearby companion.  Specifically, a low-mass companion in a $0.1$~AU orbit would impart sufficient angular momentum to spin-up the outer stellar layers before being engulfed as the star ascended the giant branch while not dramatically altering the stellar evolution.  

In this series of papers, we have demonstrated that precision interferometry at CHARA is capable of detecting the faint main-sequence companions of bright RS CVn primary stars.  We are currently processing new data for another close, bright RS CVn system with the intention to publish these results in a follow-up paper.
 
\section*{Acknowledgements}

We thank F.\ C.\ Adams, J.\ A.\ Orosz,  D.\ Pourbaix, and M.\ Rieutord for their help and comments, as well as L.\ Boyd of Fairborn Observatory for his support of the photometric observations.  The interferometric data in this paper were obtained at the CHARA Array, funded by the National Science Foundation through NSF grants AST-0908253 and AST-1211129, and by Georgia State University through the College of Arts and Sciences.  The MIRC instrument at the CHARA Array was funded by the University of Michigan.  NSF grant 1039522 from the Major Research Instrumentation Program, awarded to Tennessee State University, made extracting the secondary velocities possible.  Astronomy at Tennessee State University is supported by the state of Tennessee through its Centers of Excellence program.  The APT program was supported by NASA, NSF, Tennessee State University, and the Centers of Excellence Program.  Nordic Optical Telescope is operated on the island of La Palma jointly by Denmark, Finland, Iceland, Norway, and Sweden, in the Spanish Observatorio del Roque de los Muchachos of the Instituto de Astrofisica de Canarias.
R.M.R.\ would like to acknowledge support from the NASA Harriet G.\ Jenkins Pre-Doctoral Fellowship and a Sigma Xi Grant-in-Aid of Research.  
J.D.M.\ and R.M.R.\ acknowledge support of NSF grant AST-1108963.
This research has made use of the SIMBAD database, operated at CDS, Strasbourg, France and the Jean-Marie Mariotti Center \texttt{SearchCal} service\footnote{Available at http://www.jmmc.fr/searchcal}
co-developed by FIZEAU and LAOG/IPAG, and of CDS Astronomical Databases SIMBAD and VIZIER\footnote{Available at http://cdsweb.u-strasbg.fr/}.
 
\appendix
\section{A. AST Radial Velocities}

\begin{figure}
\begin{center}
\vspace{0cm}
\includegraphics[scale=0.8]{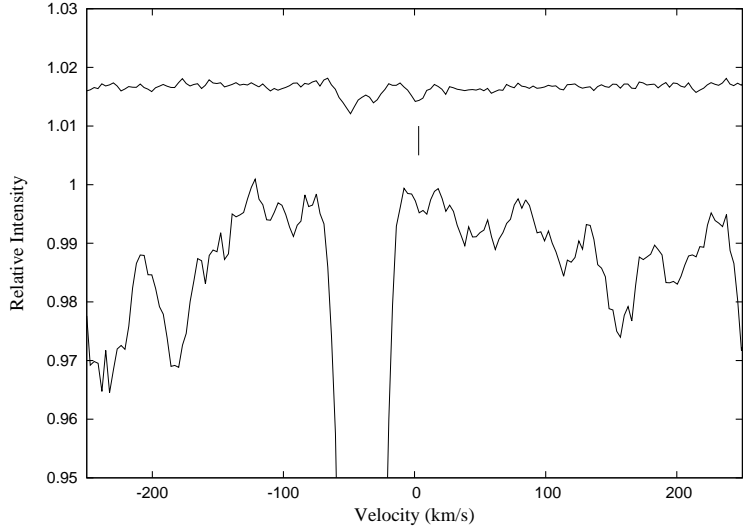}
\vspace{0cm}
\caption{From a Fairborn Observatory spectrum of $o$~Dra, the lower
solid line is the average profile of the components summed
over 168 spectral regions. The upper line, arbitrarily vertically
shifted, is the remainder after the average region around the 
primary component from all useful Fairborn spectra has been 
removed from the lower line. The position of the secondary is
indicated by a tick mark below the summed residual spectrum.
}
\label{omiDrasecspec}
\end{center}
\end{figure}

The secondary lines of $o$~Dra were not obvious in the initial 
measurements of our individual AST spectra. Thus, similar to the
method used by \citet{fek15}, we subtracted the spectrum of the 
primary, which was obtained by averaging our spectra, appropriately 
shifted so that all the primary lines from spectrum to spectrum 
were aligned. This subtraction resulted in a very weak average 
summed profile of the lines in the residual spectra corresponding 
to the secondary component (Figure \ref{omiDrasecspec}).  A Gaussian function was 
fitted to the very weak average secondary component to obtain its
radial velocity. Because of the extreme weakness of the secondary 
lines and variable signal-to-noise ratio of the various spectra, 
most of the residual AST spectra did not produce measurable 
features of the secondary.

For the primary we determined a $v \sin i$ value of $15.6 \pm 1.0$~
km s$^{-1}$, which is in good agreement with the value of $16.0 
\pm 0.5$ km s$^{-1}$ from the NOT spectra. The equivalent width 
ratio of the average component lines, A/B, which corresponds to 
the continuum intensity ratio, was measured in several AST spectra 
that had the highest signal-to-noise ratios. We find a ratio of 
128 $\pm$ 4 at a mean wavelength of about 6000~\AA, a value 
that is similar to the $V$ band flux ratio of $130$ found from 
the photometry.

The parameters from the AST double-lined spectroscopic orbit and the combined \citet{mas08}/CfA orbits are presented in Table \ref{RVonlyorb}.

\begin{deluxetable}{l c c }
\tablecaption{$o$~Dra Orbital Parameters from Radial Velocity Curves}
\tablehead{
\colhead{Parameter}         &
\colhead{\citet{mas08}/CfA} &
\colhead{AST/TSU}  
}
\startdata
orbital period,             $P_\mathrm{orb}$ (days)  & $138.45 \pm 0.03$  &  $138.46 \pm 0.01$    \\
center of mass velocity,    $\gamma$ (km s$^{-1}$)   & $-20.8 \pm 0.3$ & $-20.87 \pm 0.02$  \\
semi-amplitude, primary,    $K_\mathrm{A}$ (km s$^{-1}$) & $23.48 \pm 0.31$ & $23.37 \pm 0.03$ \\
semi-amplitude, secondary,  $K_\mathrm{B}$ (km s$^{-1}$) & & $31.86 \pm 0.42$ \\
eccentricity,               $e$               & $0.18 \pm 0.02$ & $0.154 \pm 0.001$ \\
time of periastron passage, $T$ (HJD)     & $2454982 \pm 2$ &  $2455812.8 \pm 0.2$     \\
longitude of periastron,         $\omega_\mathrm{A}$ ($^\circ$)  & $292 \pm 5$ & $290.7 \pm 0.4$ \\
$M_\mathrm{A} \sin^3 i$  ($M_\odot$)           & &   $1.34 \pm 0.04$    \\
$M_\mathrm{B} \sin^3 i$  ($M_\odot$)           & &   $0.99 \pm 0.01$   \\
$a_\mathrm{A} \sin i$ ($10^6$ km)                 & & $43.97 \pm 0.05$ \\
$a_\mathrm{B} \sin i$ ($10^6$ km)   & & $59.92 \pm 0.79$ \\
$a \sin i$ ($R_\odot$)                             &  $63.2 \pm 0.9$ &\\
mass function, $f(M) = (M_\mathrm{B} \sin i)^3/(M_\mathrm{A} + M_\mathrm{B})^2$ & $0.177 \pm 0.007$ & \\
RMS velocity residuals, $\sigma_\mathrm{A}$ (km s$^{-1}$) & 0.39 & $0.16$ \\
RMS velocity residuals, $\sigma_\mathrm{B}$ (km s$^{-1}$) & & $1.5$ 
\enddata
\label{RVonlyorb}
\end{deluxetable}

\section{B. Interferometric Observables}

\begin{figure}
\begin{center}
\vspace{0cm}
\includegraphics[angle=90,scale=.4]{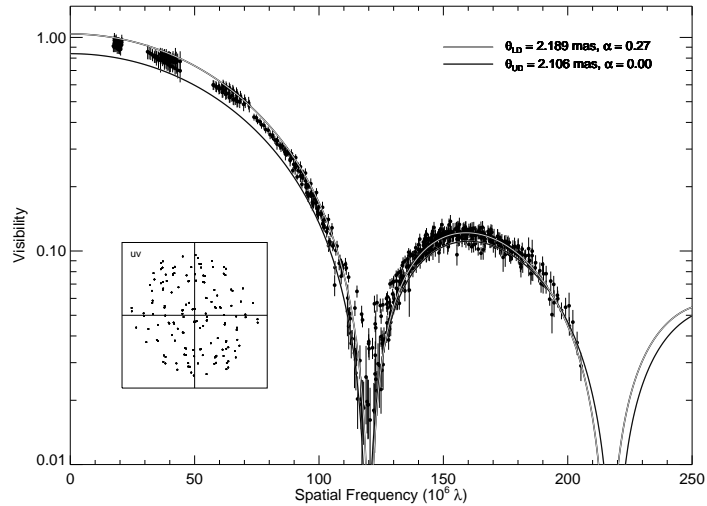}
\vspace{0cm}
\caption{Visibility curve of the 2012 Jun 18 observations of $o$~Dra with CHARA/MIRC.  The observed visibility curve is plotted in black with $1 \sigma$ error bars.  The best-fit model from fitting for the companion is overplotted in as the black line ($\theta_\mathrm{UD} = 2.106$~mas, $V(0) = 0.839$, $\alpha = 0.00$).  The white line is our limb-darkened model ($\theta_\mathrm{LD} = 2.189$~mas, $V(0) = 1.038$, $\alpha = 0.27$.  The inset is the $uv$-coverage on the night of observation.}
\label{omiDraVis}
\end{center}
\end{figure}

\begin{turnpage}
\begin{figure*}
\hspace{-0.5cm}
\vspace{0cm}
\includegraphics[angle=90,scale=0.95]{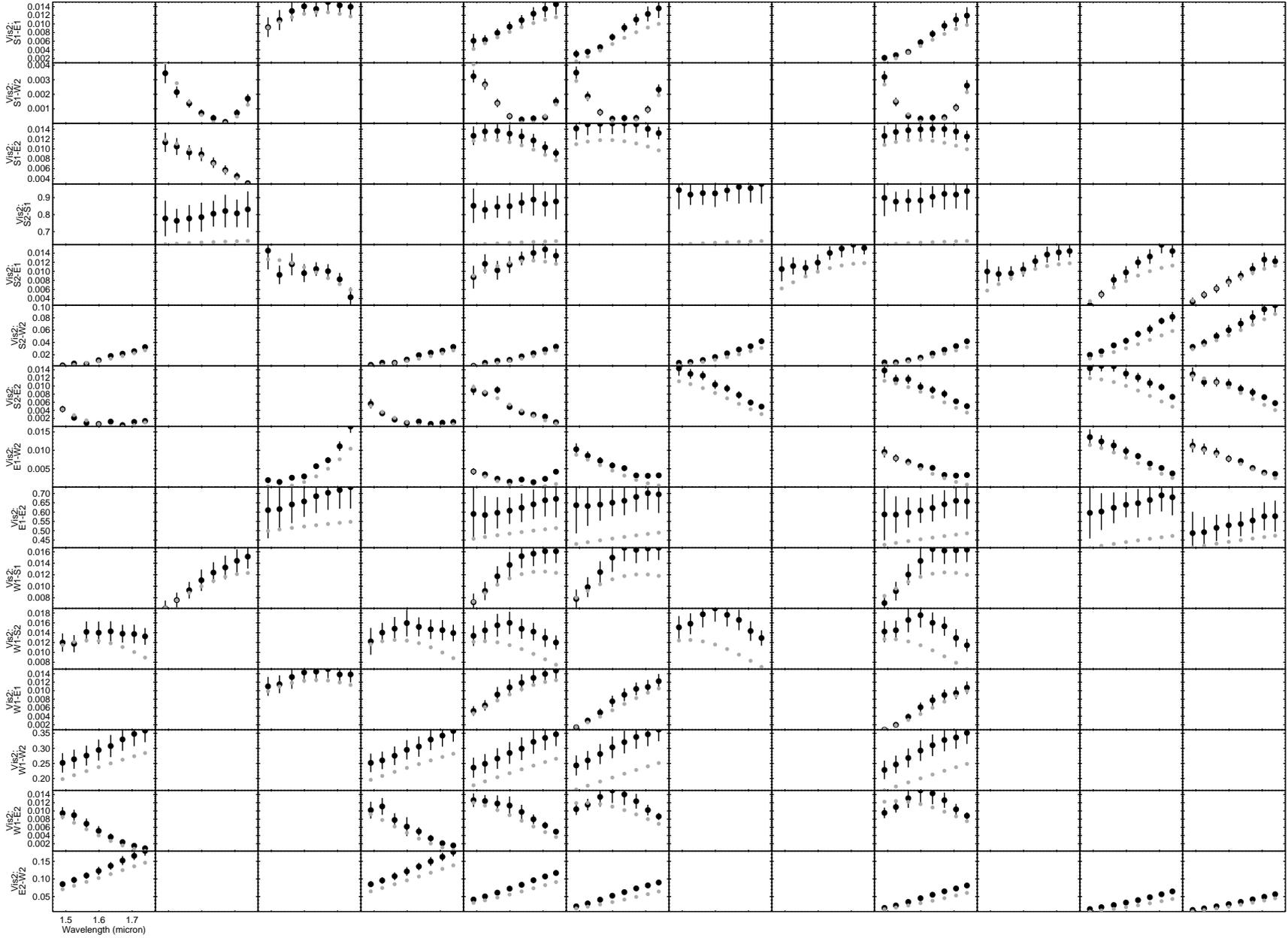}
\vspace{0cm}
\caption{Squared visibilities of the 2012 Jun 18 observations of $o$~Dra with CHARA/MIRC.  Each block of observations represents a temporal block of observations.  The data and model are plotted as in Figure \ref{omiDraVis}.}
\label{omiDraV2}
\end{figure*}
\end{turnpage}

\begin{turnpage}
\begin{figure*}
\hspace{-0.5cm}
\vspace{0cm}
\includegraphics[angle=90,scale=0.95]{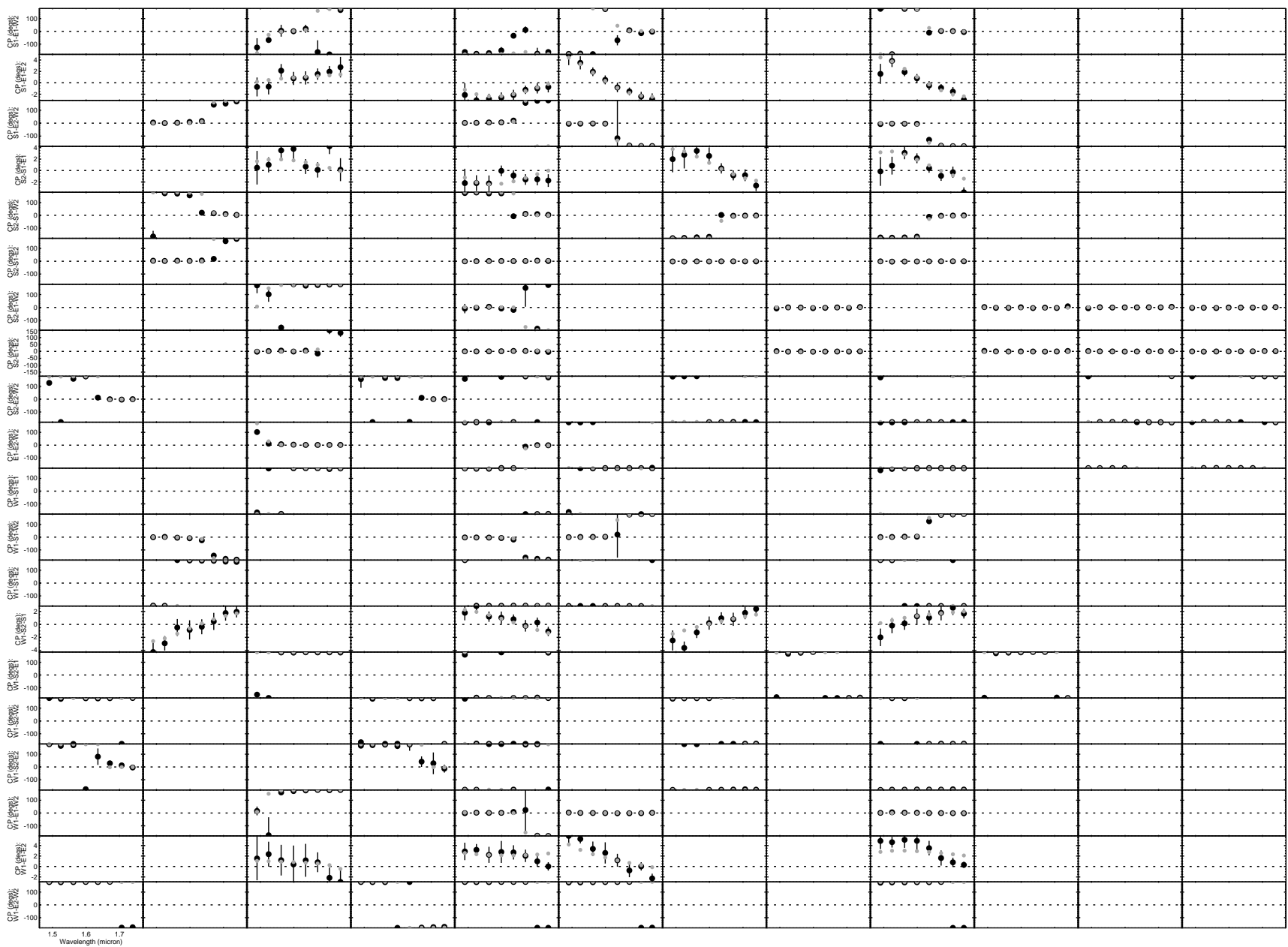}
\vspace{0cm}
\caption{Closure phases of the 2012 Jun 18 observations of $o$~Dra with CHARA/MIRC plotted as in Figure \ref{omiDraV2}.}
\label{omiDraCP}
\end{figure*}
\end{turnpage}

\begin{turnpage}
\begin{figure*}
\hspace{-0.5cm}
\vspace{0cm}
\includegraphics[angle=90,scale=0.95]{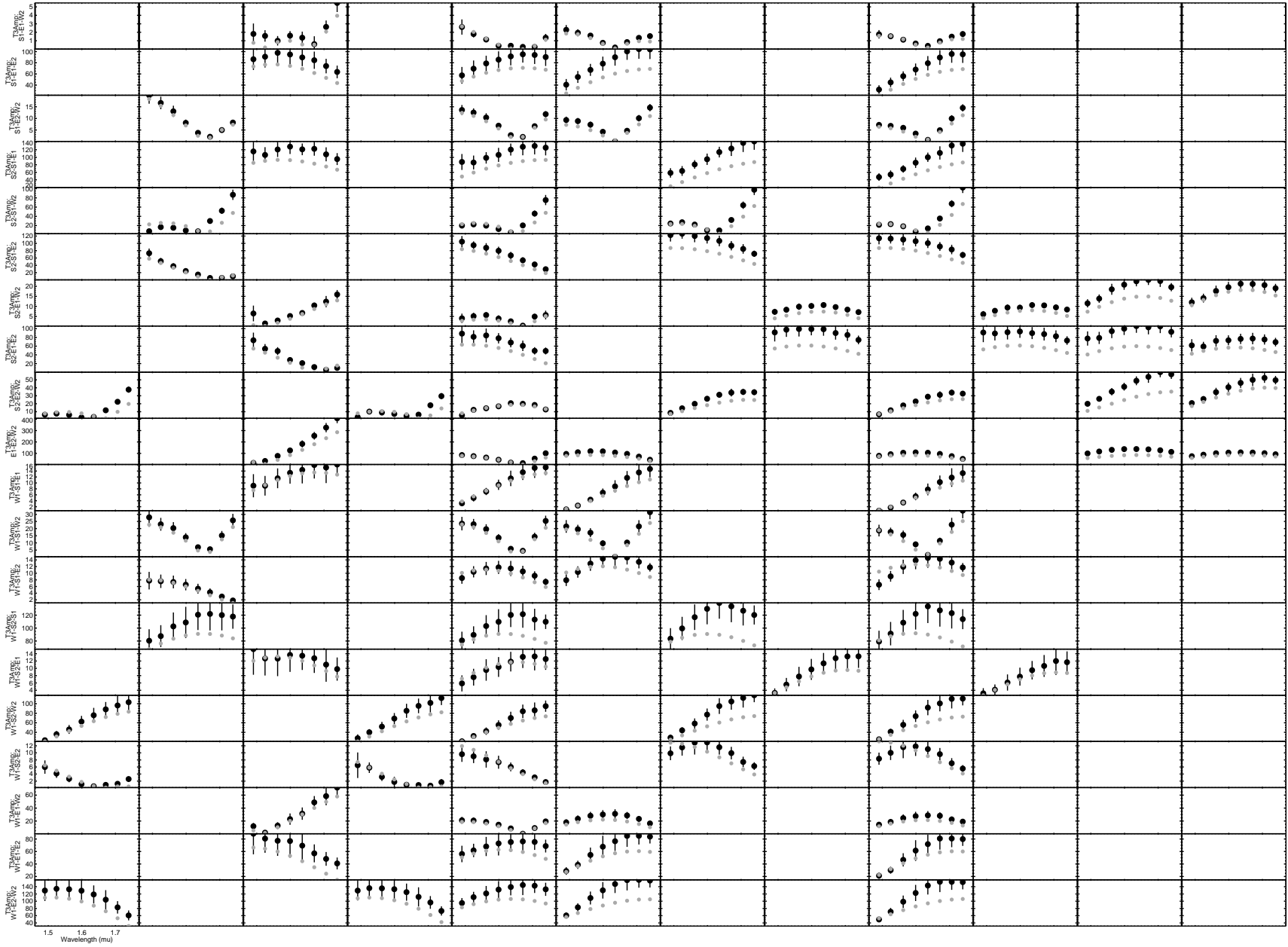}
\vspace{0cm}
\caption{Triple amplitudes (multiplied by $10^4$ for clarity) of the 2012 Jun 18 observations of $o$~Dra with CHARA/MIRC plotted as in Figure \ref{omiDraV2}.}
\label{omiDraT3}
\end{figure*}
\end{turnpage}

In Figures \ref{omiDraVis} - \ref{omiDraT3}, we present a sample comparison of the calibrated $o$~Dra data from 2012 Jun 18 and the best-fit detection of the companion from our $\chi^2$-space fit.

\end{document}